\documentclass[aps,prd,superscriptaddress,twocolumn
]{revtex4-1}
\usepackage{graphicx}
\usepackage{amsmath}
\usepackage{mathtools}
\usepackage{amsfonts}
\usepackage{bm}
\usepackage[normalem]{ulem}
\usepackage[usenames, dvipsnames]{xcolor}

\usepackage{hyperref}
\hypersetup{colorlinks=true, linkcolor=blue, citecolor=blue, urlcolor=blue}

\begin{document}

\title{Infinite-layer fluoro-nickelates as $d^9$ model   materials}

\author{F. Bernardini}
\affiliation{Dipartimento di Fisica, Universit\`a di Cagliari, IT-09042 Monserrato, Italy}

\author{V. Olevano}
\affiliation{Institut N\'eel, CNRS \& UGA, 38042 Grenoble, France}

\author{X. Blase}
\affiliation{Institut N\'eel, CNRS \& UGA, 38042 Grenoble, France}

\author{A. Cano}
\affiliation{Institut N\'eel, CNRS \& UGA, 38042 Grenoble, France}
\date{\today}

\begin{abstract}

We study theoretically the fluoro-nickelate series $A$NiF$_2$ ($A=$ Li, Na, K, Rb, Cs) in the tetragonal $P4/mmm$ infinite-layer structure. 
We use density functional theory to determine the structural parameters and the electronic band structure of these unprecedented compounds. 
Thus, we predict these materials as model $d^9$ systems where the Ni$^{1+}$ oxidation is realized and the low-energy physics is completely determined by the Ni-3$d$ bands only. 
Fluoro-nickelates of this class thus offer an ideal platform for the study of intriguing physics that emerges out of the special $d^9$ electronic configuration, notably high-temperature unconventional superconductivity.
\end{abstract}

\maketitle

The recent discovery of superconductivity in Sr-doped NdNiO$_2$ thin films \cite{li-nature19} 
has attracted a remarkable research attention (see e.g. \cite{botana19,sakakibara19,hirsch19,jiang19,wu19,nomura19,bernardini19,hepting19,zhang19effective,zhang19,zhang19a,pickett19,hu19,ryee19,hirayama19}). 
In these infinite-layer nickelates the Ni atom takes a nominal ${1+}$ valence with a 3$d^9$ electronic configuration that leads to a 3$d_{x^2-y^2}$ band crossing the Fermi level. This configuration is rare in nature. At the same time, it is one of the fingerprints of high-$T_c$ superconducting cuprates. That is, in fact, the basis of an intriguing analogy put forward some time ago \cite{anisimov99,pickett04}, which now receives a renewed interest and motivates the search for novel $d^9$ materials \cite{hirayama19}.  
Such an analogy, however, is fundamentally limited among other factors by the presence of extra $A$-atom bands in the $A$NiO$_2$ nickelates that also intersect the Fermi level and play a special role \cite{hepting19,zhang19effective,zhang19,zhang19a,pickett19}.  

\begin{figure}[b!]

\includegraphics[width=0.25\textwidth]{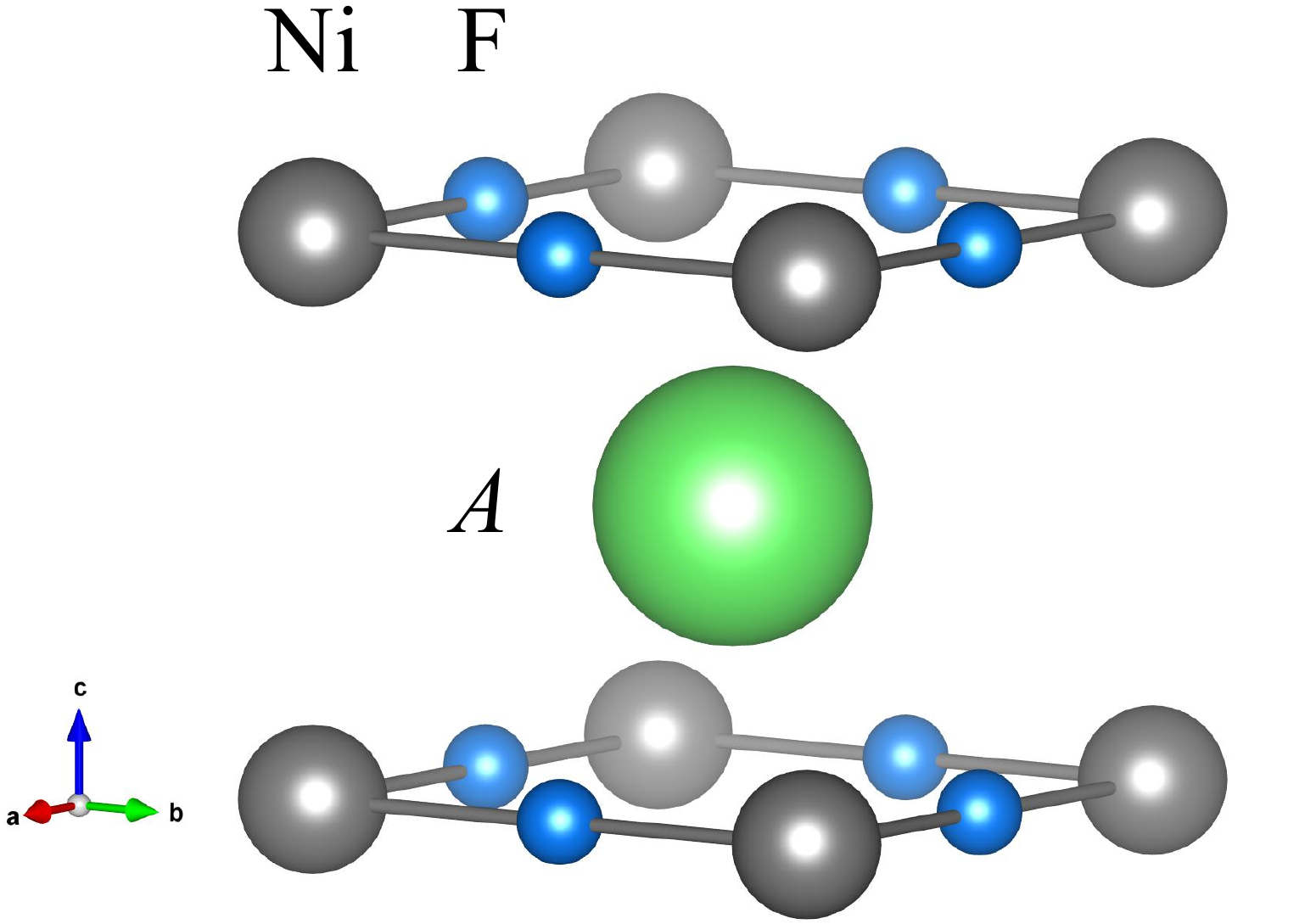}
\includegraphics[width=0.15\textwidth]{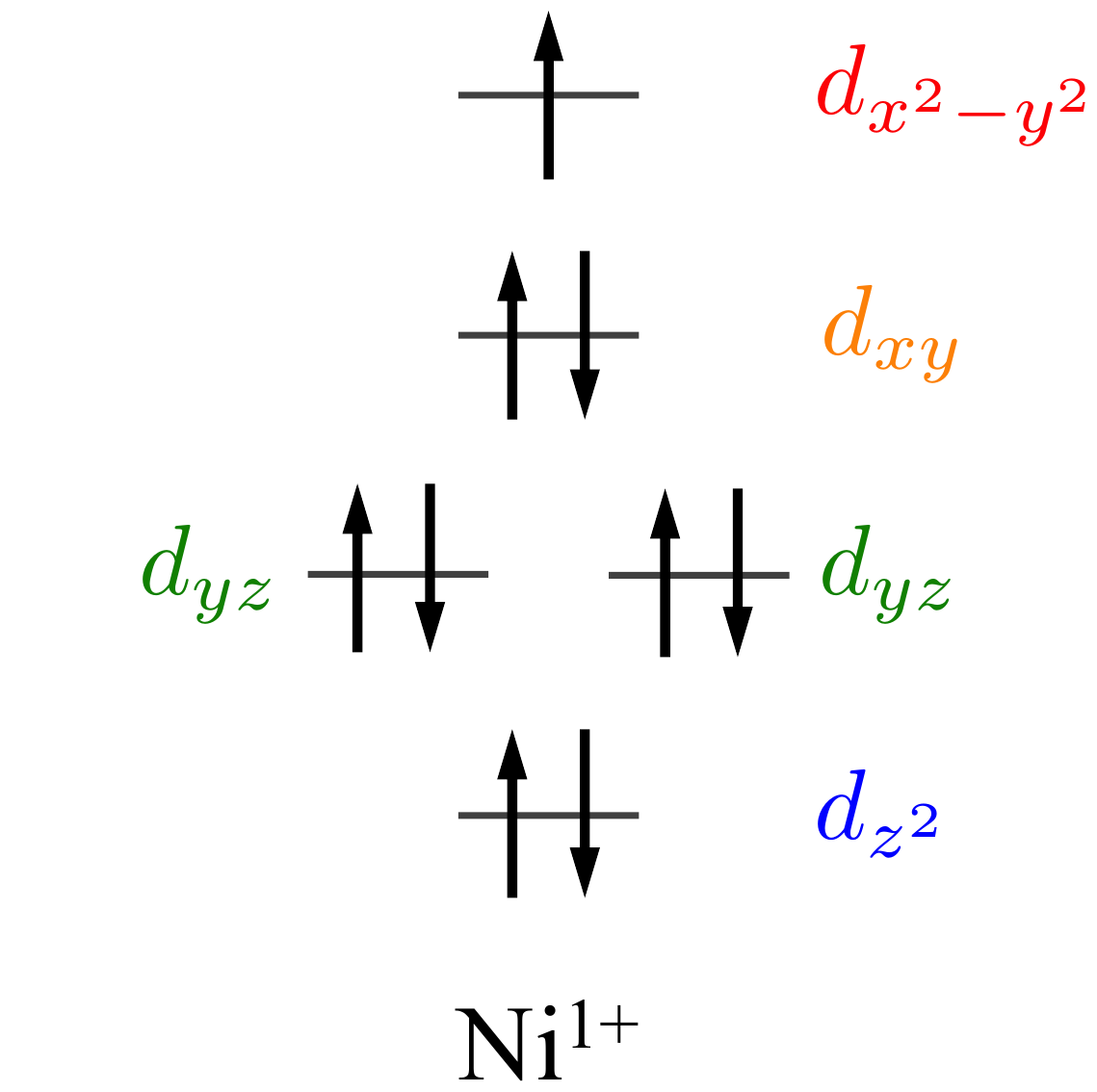}
  \caption{Ball-and-stick model of the $A$NiF$_2$ fluoro-nickelates in the $P4/mmm$ space group (left) and 3$d^9$ electronic configuration of the Ni$^{1+}$ under the corresponding crystal field (right). }

\label{fig:1}
\end{figure}

Here, we investigate theoretically the novel infinite-layer fluoro-nickelate series $A$NiF$_2$, with $A=$ Li, Na, K, Rb, Cs (see Fig. \ref{fig:1}). These materials can in principle be synthesized from existing fluoro-perovskite precursors through a topotactic reduction process similar to that used to obtain their oxide counterparts \cite{crespin83,hayward99,crespin05,li-nature19}. In such a $A$NiF$_2$ series the Ni atom is combined with monovalent elements so that the Ni$^{1+}$ oxidation state can be naturally expected. Besides, no extra $A$-atom band is expected near the Fermi level and a sizable separation between Ni-3$d$ and F-2$p$ bands ---that is, a large charge transfer energy--- can be anticipated given the fact that F is the most electronegative element.  
In the following, we report first-principles calculations based of density functional theory (DFT) that confirm these expectations. Specifically, we compute the lattice parameters and the electronic band structure of the $A$NiF$_2$ series ($A=$ Li, Na, K, Rb, Cs). In addition, we study the effect of external pressure and doping together with the tendency to magnetism of these new compounds.
We illustrate our results using KNiF$_2$ as reference system and report on the rest of fluoro-nickelates in the Appendix. 

\section{Methods}

We performed DFT calculations to study the crystal structure and electronic band structure of the $A$NiF$_2$ fluoro-nickelates. We used the VASP code \cite{VASP} to determine the lattice parameters of these systems in the $P4/mmm$ structure, with the PAW pseudo-potentials~\cite{PAW} and the PBE~\cite{PBE} exchange and correlation functional. 
A plane-wave cutoff of 960 eV and a Monkhorst-Pack $6\times6\times6$ mesh with a 0.2 eV smearing was used for the calculations. A test performed on the KNiF$_2$ with a more accurate mesh showed that the structural parameters were converged within 0.002 \AA. 
The corresponding band structures were computed using the full-potential linear augmented plane-wave (FLAPW) method as implemented in the {\sc{WIEN2k}} package \cite{Wien2k} with the LDA exchange and correlation functional \cite{LDA}. We choose muffin-tin radii of 2.5 a.u., 2.1 a.u., and 1.7 a.u. for the K (Rb, Cs), Ni, and F atoms respectively and a plane-wave cutoff $R_{\rm mt}K_{\rm max}=7.0$. The non-magnetic calculations were performed using the primitive (1-formula-unit) cell of the chemical structure and a 
Monkhorst-Pack mesh of $12\times12\times13$ $k$-points in the irreducible Brillouin zone.
The maximally localized Wannier functions (MLWFs) \cite{MLWF} were calculated by interfacing the {\sc{WANNIER90}} package \cite{Wannier90} to {\sc{WIEN2k}} using the {\sc{WIEN2WANNIER}} code \cite{Wien2wannier}. 

The magnetic solutions were obtained using $\sqrt{2}\times\sqrt{2}\times 2$ supercells with equivalent $k$-point meshes for the following configurations (see e.g. \cite{bousquet16} for the definitions): ferromagnetic (FM), $A$-type antiferromagnetic (AFM) with spins pointing in opposite directions along $c$ (AFM order of FM planes), $C$-type AFM with spins pointing in opposite directions in consecutive lines (AFM order of FM chains), $G$-type AFM with nearest–neighboring spins pointing in opposite directions (``full" AFM ordering). In addition we also considered the so-called $E$-type AFM order using an adapted supercell. 
The effect of charge carrier doping was studied using a charged-cell approach with a compensating background.  

\section{Results}

\subsection{Crystal structure}

The calculated lattice parameters of the fluoro-nickelate series $A$NiF$_2$ in the $P4/mmm$ structure are summarized in Table~\ref{tab:struct}. The computed $a$ parameter is comparable to that of the oxide nickelates (see e.g. \cite{li-nature19}). However, there is a substantial change in the $c$ parameter as a function of the radius of the monovalent $A$ atomacross the series. This yields a monotonous increase both in the $c/a$ ratio and in the volume of the unit cell with increasing such a  radius. Interestingly the lattice parameters become $a < c$ for Rb and Cs, unlike in the oxide counterparts. 
This behavior suggests that the binding of $A$ atom with the NiF$_2$ layer is due to the electrostatic interactions rather than to covalent bonding. 

Table~\ref{tab:struct} also summarizes the lowest-energy phonons at $\Gamma$ and $R$ points calculated in the $P4/mmm$ infinite-layer structure. Structural instabilities in the corresponding perovskites can be identified from these phonons \cite{garcia-castro-prb13}. 
We find rather soft modes for LiNiF$_2$ and NaNiF$_2$, which signals the vicinity to a structural quantum critical point in these two systems. 
This is an interesting possibility whose study, however, is beyond the scope of the present work. 
In fact, even if such a $P4/mmm$ structure may be unstable for some fluoro-nickelates in their bulk form, the calculated structures can be stabilized in thin films by means of epitaxial strain. 

\begin{table}[b!]
 \begin{tabular}{c c c c c c }
\hline \hline
                    &  LiNiF$_2$  &  NaNiF$_2$  & KNiF$_2$  & RbNiF$_2$  & CsNiF$_2$  \\
\hline
$a$ (\AA)           &  3.956      &  4.008      & 4.040     &  4.046     & 4.057 \\
$c$   (\AA )        &  2.815      &  3.058      & 3.832     &  4.224     & 4.645 \\
\hline 
$\hbar\omega_\Gamma$ (meV) &  - &  16 & 16 & 12 & 10 \\
$\hbar\omega_R$ (meV) &  - &   - & 13 & 10 &  8 \\
\hline \hline
\end{tabular}
\caption{Calculated lattice parameters of $A$NiF$_2$ fluoro-nickelates in the tetragonal $P4/mmm$ structure (non-magnetic state), and lowest energy of the $\Gamma$-point (optical) and $R$-point phonons. The symbol - indicates that a soft phonon of very low energy is obtained in the calculation.}
\label{tab:struct}
\end{table}

\subsection{Band structure}
\begin{figure}[t!]
\includegraphics[width=0.475\textwidth]{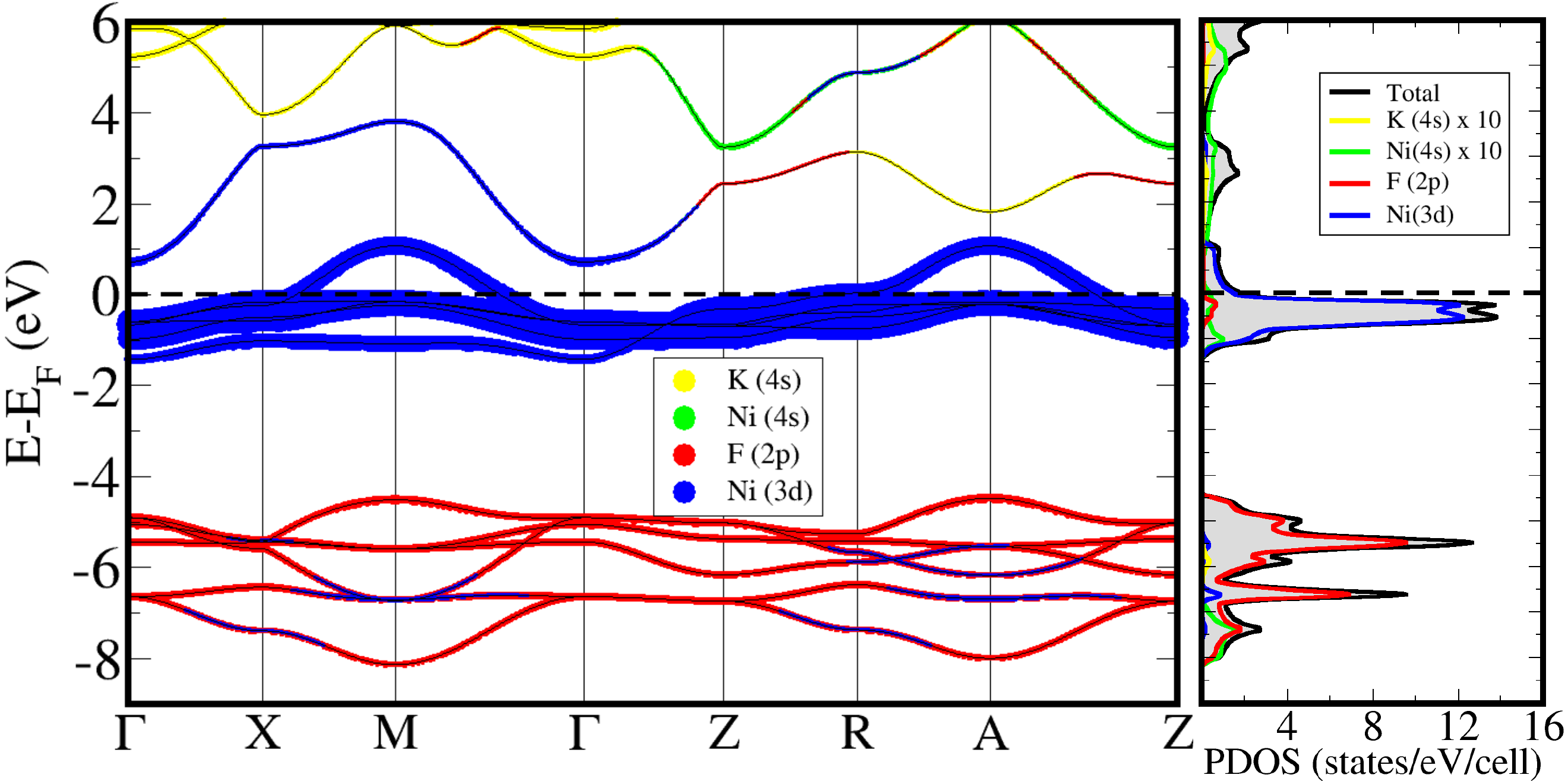}
  \caption{
  Calculated band structure of KNiF$_2$ in the non-magnetic phase. The main orbital character of the bands is emphasized by the different colors as indicated in the legend together with the width of the lines (fatband plot). Such a fatband plot can produce some ``visual artifacts'' for the less localized bands (thinner lines) that can be resolved by inspecting the projected density of states (PDOS). The K-4$s$ and Ni-4$s$ components of the PDOS are scaled by a factor of 10. } 
\label{fig:NM_WIDE}
\end{figure}

Fig.~\ref{fig:NM_WIDE} shows the band structure of KNiF$_2$, which is a prototype for the $A$NiF$_2$ series. The bands associated to the F-2$p$ orbitals are well separated from the Ni-3$d$ ones by an ionic gap of $\sim$ 3 eV and lie below $\sim -5$~eV.
This is in sharp contrast to the oxide nickelates and even more to the analog cuprates. In the former such an ionic gap nearly closes, while in the latter O-2$p$ and Cu-3$d$ bands actually overlap. According to this feature, the fluoro-nickelates can provide actual materials in which the Mott insulator limit in the Zaanen-Sawatzky-Allen scheme can be realized in a clean way \cite{jiang19}.

When it comes to the Ni-3$d$ bands, they form a compact set near the Fermi level. Their bandwidth ranges from $\sim $\,2.5 eV to $\sim $\,0.5 eV, and therefore are visibly flatter that in the oxide nickelates. Importantly, the Ni-3$d$ bands are well separated from the next set of bands higher in energy, none of which intersect the Fermi level. 
This higher-energy set is much less localized and consequently its orbital character is more hybrid. According to the DOS, it is mainly associated to K-4$s$ and Ni-4$s$ orbitals with a non-negligible contribution from the Ni-3$d_{z^2}$ ones. 

\begin{figure*}[t!]
\includegraphics[width=.9\textwidth]{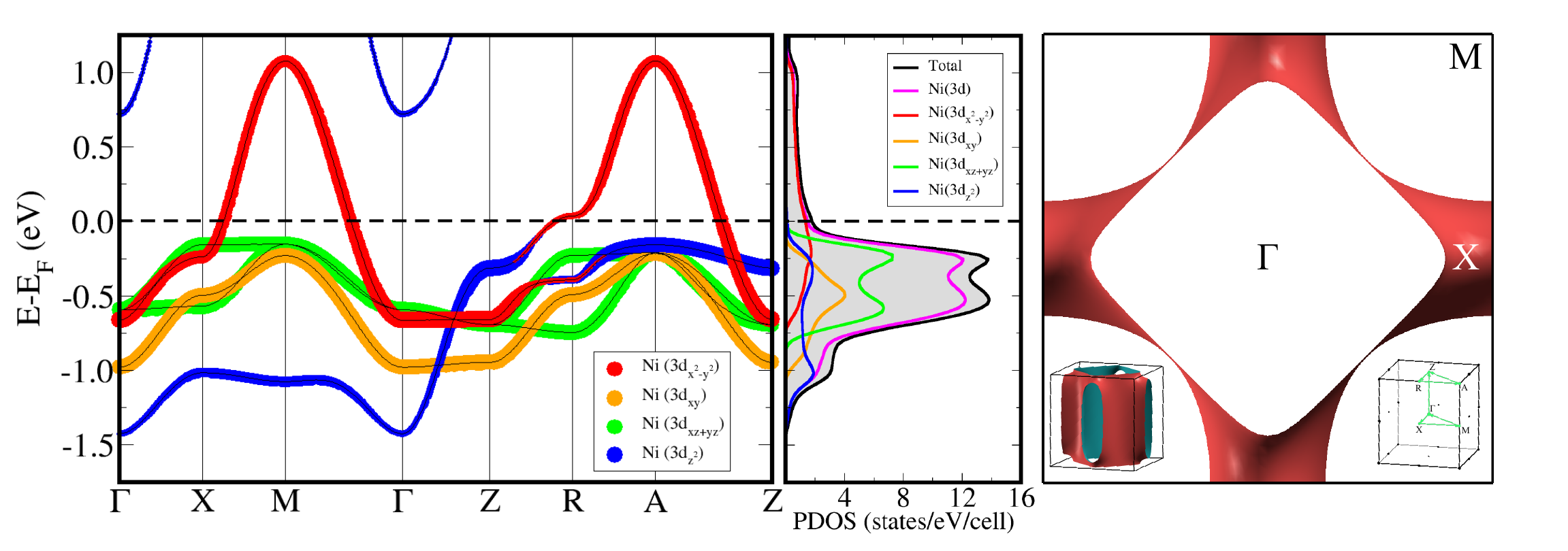}
  \caption{(Left) Zoom of the electronic band structure of KNiF$_2$ near the Fermi energy in a ``fatband" plot, (middle) orbital resolved density of states and (right) top view of Fermi surface (the insets show a perspective view of the Fermi surface and the Brillouin zone with the high-symmetry points and the $k$-path used to plot the band structure).}  
\label{fig:NM}
\end{figure*}

\begin{figure*}[t!]
\includegraphics[width=.9\textwidth]{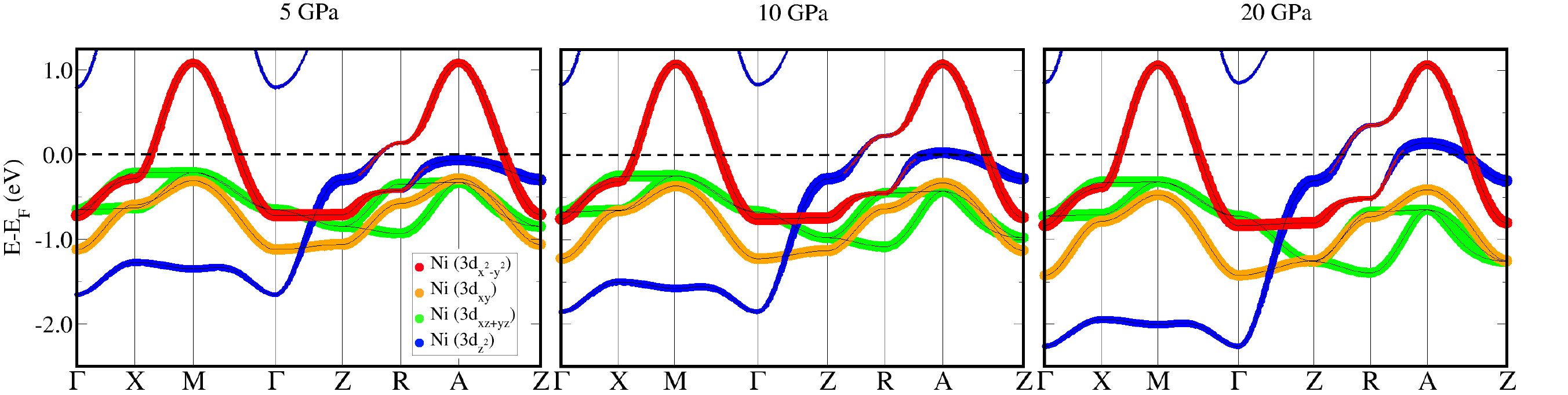}
  \caption{Orbital resolved band structure of KNiF$_2$ under hydrostatic pressure.} 
\label{fig:press}
\end{figure*}

Fig.~\ref{fig:NM} shows the Ni-3$d$ bands resolved according to their orbital character in more detail.
As we see, only the Ni-3$d$ bands are relevant for the low-energy physics and the 3$d_{x^2-y^2}$ character dominates the crossing with the Fermi level. This is as expected according to the $d^9$ configuration of the Ni$^{1+}$ with the nominally half-filled 3$d_{x^2-y^2}$ orbital (see Fig. \ref{fig:1}). However, the crossing with the Fermi level actually involves two different bands that swap their character and become 3$d_{z^2}$ below the Fermi energy at $M$ and $A$. This feature traces back to an avoided crossing, which can be best appreciated by inspecting the band structure of the whole $A$NiF$_2$ series (Figs. \ref{fig:NM} and \ref{fig:series}). 
In fact, the Fermi surface wraps around $M$ at $k_z=0$ while it wraps around $Z$ at $k_z=\pi/c$. Thus, despite its apparent single-band two-dimensional character, the Fermi surface is in reality a more complex three-dimensional object. The three-dimensional character of the Fermi surface has also been pointed out for LaNiO$_2$ \cite{pickett04}. However, to the best of our knowledge, its ``multiband" nature related to the avoided crossing with orbital character exchange has been overlooked so far (even if it can be deduced from Fig. 3 in \cite{pickett04}). A closely related hybridization of this type has been suggested to play a key role in cuprates \cite{sakakibara12}. 
Since this feature emerges in the $A$NiF$_2$ fluoro-nickelates in a very clean and prominent way, these systems can indeed be of great help to clarify the actual situation.

\begin{table}[b!]
 \begin{tabular}{c c c c c c}
\hline \hline
                    &  0 GPa         &    5 GPa       &  10 GPa      & 20 GPa     \\
\hline
$a$ (\AA)           &  4.040      &  4.004      & 3.986     &  3.957     \\
$c$   (\AA )        &  3.832      &  3.491      & 3.265     &  2.978     \\
\hline \hline
\end{tabular}
\caption{Structural parameters of KNiF$_2$ as a function of pressure in the tetragonal $P4/mmm$ structure (non-magnetic state).}
\label{tab:press}
\end{table}

We also studied the changes in the band structure that can be obtained by the application of hydrostatic pressure (see Fig. \ref{fig:press}). The corresponding lattice parameters are listed in Table~\ref{tab:press}. We see that $a$ remains essentially constant while $c$ decreases significantly under pressure. 
This reduction increases the overlap between the 3$d_{z^2}$ orbitals of neighbouring Ni atoms, and further yields an enhanced dispersion of the 3$d_{z^2}$ band between $\Gamma$ and $Z$ (see Fig. \ref{fig:press}).     
Beyond that, pressure effectively acts as a sort of hole doping in the sense that it produces a shift upwards of a portion of the bands near the Fermi level. This enhances the $d_{z^2}$ orbital content of the Fermi surface and eventually gives rise to
an additional hole pocket at $A$ (where the corresponding band is rather flat and therefore can yield a sizable van Hove feature). Interestingly, pressure turns out to be an effective control parameter for tuning the avoided crossing with band-character inversion to the Fermi level.

\subsection{Wannier functions}
In order to gain further insight into the electronic structure of the fluoro-nickelates and compare them with their oxide counterparts, we determined the Wannier orbitals of KNiF$_2$ in the non-magnetic state. The reduced dispersion and compact nature of both Ni-3$d$ and F-2$p$ band sets simplifies the calculation of the MLWFs and makes this calculation possible without disentanglement procedure. 
We used the lowest eleven bands shown in Fig.~\ref{fig:NM_WIDE} to construct eleven MLWFs. These MLWFs are associated to five Ni-3$d$ and six F-2$p$ orbitals, three for each F atom. The spatial spread of this functions is small ($< 1.2$ \AA$^2$). The on-site energies and hoppings obtained from the Wannier fits are shown in Table~\ref{tab:wann}. Compared to the values computed for LaNiO$_2$ in \cite{botana19,jiang2019electronic}, we find that there is a swap between $d_{x^2-y^2}$ and $d_{xz/yz}$  on-site energies toghether with a reduction of difference between the $d_{x^2-y^2}$ and the $d_{z^2}$ ones to $0.1$ eV.
On the other hand, the diference between the Ni-$d_{x^2-y^2}$ and the F-$p_{x}$ energies increases to $\sim$6 eV. This is much larger than in the case of LaNiO$_2$ and LaCuO$_2$ \cite{botana19}, as expected from the enhanced electronegativity of the F. This naturally leads to a more localized $d_{x^2-y^2}$ Wannier function, which will be relevant for the problem of polaron formation. In any case, enhanced electronic correlations be naturally expected due to such an increased localization. 
The Ni-3$d$ and F-2$p$ hopping parameters are 20\% smaller than in LaNiO$2$, but otherwise their relative strength is similar. This indicates that the overlap between Ni and F centered orbitals is as in the oxide counterparts. Finally, the hopping between neighboring F-centered orbitals are very small, confirming the much more localized nature of the 2$p$ orbitals of F with respect to O.  

\begin{table}[t!]
 \begin{tabular}{p{5em} r p{4em} p{5em} r}
\hline \hline
\multicolumn{5}{l}{Wannier on-site energies (eV)} \\
\hline
$d_{xy}$       &    $-$0.84   && &  \\ 
$d_{xz/yz}$    &    $-$0.59   && $p_{x}$ ($\sigma$)   &     $-$6.51  \\  
$d_{x^2-y^2}$  &    $-$0.65   && $p_{y}$ ($\pi$)      &     $-$5.29  \\
$d_{z^2}$      &    $-$0.75   && $p_{z}$ ($\pi$)      &     $-$5.37  \\
\hline 
\multicolumn{5}{l}{Wannier hoppings (eV)}   \\
\hline
$d_{xy}$\,-\,$p_{y}$  &   0.61 &&& \\
$d_{xz}$\,-\,$p_{z}$  &   0.64 &&& \\
$d_{x^2-y^2}$\,-\,$p_{x}$  &  $-$1.05 &&&\\
\hline \hline
\end{tabular}
\caption{Calculated on-site energies and hoppings for KNiF$_2$ derived from Ni- and F-centered maximally localized Wannier functions (MLWF). The values associated to the F MLWFs are degenerate and the label $\sigma$ ($\pi$) indicates 2$p$ orbitals with parallel (perpendicular) orientation to the Ni - F bond.}
\label{tab:wann}
\end{table}

\subsection{Spin-polarized calculations}

\begin{figure}[t!]
\includegraphics[width=0.45\textwidth]{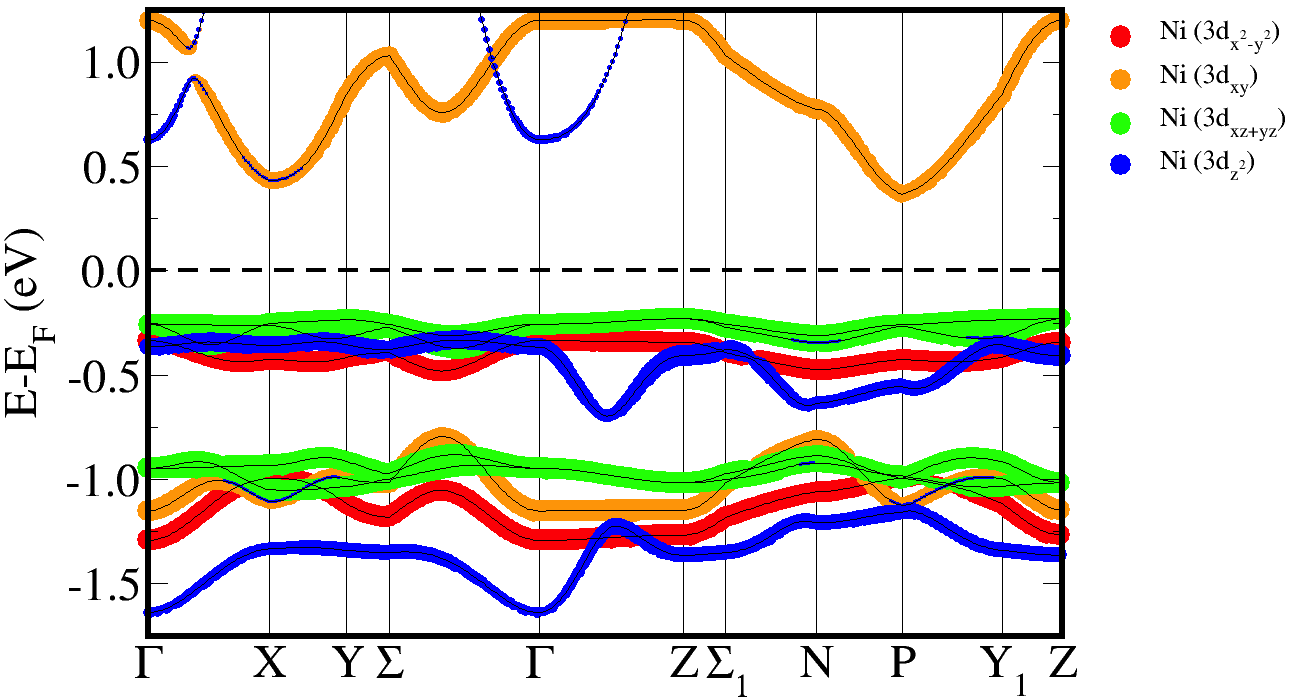}
  \caption{Orbital resolved band struture of KNiF$_2$ in the $G$-type AFM configuration.}  
\label{fig:G}
\end{figure}

\begin{figure*}[h!]
\includegraphics[width=\textwidth]{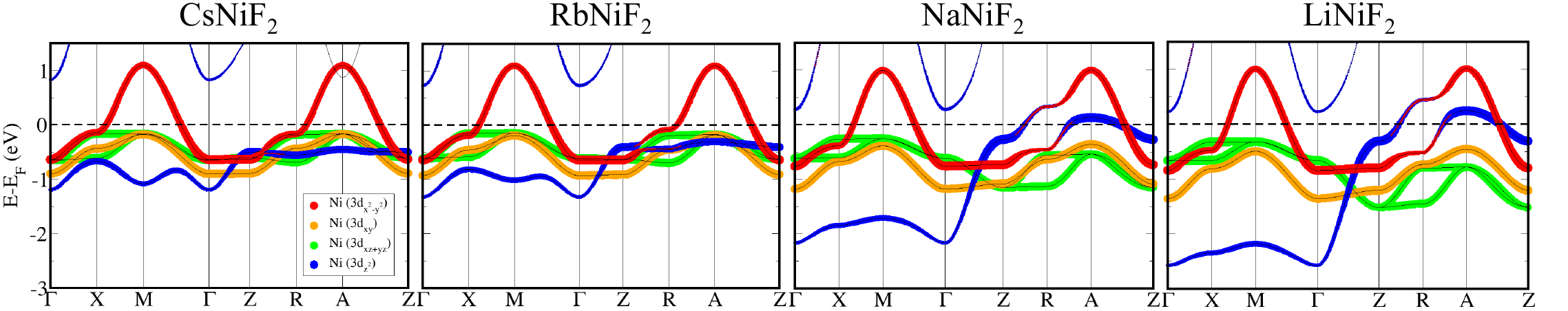}
  \caption{Fatband plot of the electronic band structure of the $A$NiF$_2$ series. The different colors and thickness emphasize the different Ni-3$d$ orbital character of the bands near the Fermi energy. These fluoro-nickelates display an avoided crossing along the $Z$-$R$ line with 3$d_{x^2-y^2}$ - 3$d_{z^2}$ exchange that evolves across the series.}  
\label{fig:series}
\end{figure*}

Next, we performed spin-polarized calculations to investigate the tendency to magnetism in the fluoro-nickelates. 
Table \ref{tab:mag} summarizes the relative energy of the magnetically ordered states with respect the non-magnetic solution in KNiF$_2$. 
The lowest energy is obtained for a $G$-type AFM state with a Ni magnetic moment of $0.87\mu_B$, which is essentially degenerate with a $E$-type AFM one. 
Given the relatively large difference between the FM magnetization energy and the $G$-type AMF one, this suggests some preference for AFM spin alignments (note that all nearest-neighbouring spins are anti-paralell in the $G$-type AFM state). In any case, these magnetization energies are much larger than that of the magnetic solutions in LaNiO$_2$. 
Indicentally, we note that such an enhanced tendency to magnetism yields an insulating $G$-type AFM state with remarkably flat bands just below the Fermi energy (see Fig. \ref{fig:G}). 

Hydrostatic pressure tends to suppress the $G$-type AFM state. Its magnetization energy becomes $-150$, $-138$ and $-112$ meV/Ni at $5$, $10$ and $20$ GPa respectively in our calculations. At the same time, the magnetisation energy of the FM state is $-14$ meV/Ni at 20 GPa. This trend suggests that pressure does not have a dramatic influence on the tendency to magnetism in KNiF$_2$. This should not be very surprising since the Ni-3$d$ electrons are very localized in the $G$-type AFM state, as can be seen from the rather flat bands that are obtained in that case (see Fig.~\ref{fig:G}). 

\begin{table}[b!]
 \begin{tabular}{l c c c c c }
\hline \hline

undoped &  FM & $A$ & $C$ & $E$ & $G$  \\
\hline
 $\Delta E$ (meV/Ni) &  $-$17 &  $-$75 & $-$159 &  $-$162 & $-$163 \\
$\mu_{\rm Ni}$ ($\mu_B$) &  0.96 & 0.99 & 0.84 &    0.90   & 0.87 \\
\hline 
0.2 h$^+$/Ni                
\\ \hline
$\Delta E$ (meV/Ni)      &  $-$54 & $-$106 & $-$187 &  $-$169         & $-$160 \\
$\mu_{\rm Ni}$ ($\mu_B$) &  1.08  & 1.10   & 1.03   &   1.05  & 1.00  \\ \hline
0.2 e$^-$/Ni                
\\ \hline
$\Delta E$ (meV/Ni)      &  - &  $-$22 & $-$70 &  $-$74          & $-$73 
\\
$\mu_{\rm Ni}$ ($\mu_B$) &  - &  0.81  & 0.67  &    0.70     & 0.69 \\ \hline
0.4 e$^-$/Ni                
\\ \hline
$\Delta E$ (meV/Ni)      &  - &  $>-$1  & $-$23 &  $-$41          & $-$24 \\
$\mu_{\rm Ni}$ ($\mu_B$) &  - &  0.11  & 0.52  &    0.50            & 0.53  \\ \hline \hline
\end{tabular}
\caption{Calculated energy difference with respect to the non-magnetic state and Ni magnetic moment for different magnetic configurations in KNiF$_2$ and different values of hole and electron doping.} 
\label{tab:mag}
\end{table}

Finally, we also studied the influence of charge carrier doping on the tendency to magnetism of KNiF$_2$. The results are sumarized in Table~\ref{tab:mag}.
We find that hole doping makes the magnetic solutions even more stable, which can be anticipated for the FM state in particular from the corresponding increase in the DOS at the Fermi level (see e.g. \cite{arribi2018}).
In fact, hole doping favors the parallel alignment of the spins with respect to the anti-parallel one. Specifically, all the orders displaying FM nearest-neighbouring spins are favored while the $G$-type AFM order, in which all these spins are AFM, is slightly suppressed. 
Electron doping has a reverse and much stronger influence in fact. In this case, all the magnetic solutions are suppressed with the FM one already absent at 0.2 e$^-$/Ni and the rest of configurations nearly degenerate in energy. A further increase to 0.4 e$^-$/Ni eventually favors the $E$-type AFM state over the $G$-type one. This reveals an intriguing magnetic softness, and suggests that electron doping can be used as a control parameter to tune and eventually suppress the tendency to magnetism in the fluoro-nickelates.

\section{Discussion}

Our results demonstrate that the infinite-layer fluoro-nickelates $A$NiF$_2$ can serve as model materials for the study of the physics associated to the $d^9$ electronic configuration. This special configuration emerges in these systems in a very clean and ideal way. Namely, the Ni-3$d$ bands
are remarkably decoupled from the rest and completely determine the low-energy electronic properties of these systems. In this sense, we anticipate similar features for $A=$ Mg, Ca, and Sr. What are the overall physical properties that emerge from such a textbook situation is a very interesting question in itself. In addition, the experimental realization of these compounds will offer an unprecedented playground to test, in particular, some important ideas that have been put forward 
in relation to the material dependence of the superconducting $T_c$ in the cuprates. On one hand, the fluoro-nickelates $A$NiF$_2$ naturally display a much larger charge-transfer energy compared to copper and nickel oxide counterparts. Consequently, electronic correlations can be expected to transform these systems into ideal Mott insulators ---rather than charge-transfer insulators--- which is believed to have a tight link to $T_c$ in cuprates. This link, however, is somehow challenged by the superconducting nickelates, which can therefore be further examined in fluoro-nickelate materials.   
On the other hand, at the DFT level, the fluoro-nickelates $A$NiF$_2$ display a Fermi surface whose topology matches that of the superconducting cuprates and is equally dominated by $d_{x^2 - y^2}$ orbital contributions. At the same time, these orbitals display an intriguing interplay with the $d_{z^2}$ ones that is important for the overall band-structure picture and the eventual Cooper pairing. In fact, this interplay emerges in the fluoro-nickelates in the most ideal and neat way, and it can be tuned across the $A$NiF$_2$ series as well as by means of pressure or charge doping. Such an interplay has also been linked to the superconducting $T_c$ \cite{sakakibara12}, so that the fluoro-nickelates can be of great help to further elucidate this point. Thus, we expect that our findings will motivate further experimental work focused on novel fluoro-nickelates as well as the search for additional $d^9$ materials.

\bibliography{fluoro-nickelates.bib}

\begin{thebibliography}{34}%
\makeatletter
\providecommand \@ifxundefined [1]{%
 \@ifx{#1\undefined}
}%
\providecommand \@ifnum [1]{%
 \ifnum #1\expandafter \@firstoftwo
 \else \expandafter \@secondoftwo
 \fi
}%
\providecommand \@ifx [1]{%
 \ifx #1\expandafter \@firstoftwo
 \else \expandafter \@secondoftwo
 \fi
}%
\providecommand \natexlab [1]{#1}%
\providecommand \enquote  [1]{``#1''}%
\providecommand \bibnamefont  [1]{#1}%
\providecommand \bibfnamefont [1]{#1}%
\providecommand \citenamefont [1]{#1}%
\providecommand \href@noop [0]{\@secondoftwo}%
\providecommand \href [0]{\begingroup \@sanitize@url \@href}%
\providecommand \@href[1]{\@@startlink{#1}\@@href}%
\providecommand \@@href[1]{\endgroup#1\@@endlink}%
\providecommand \@sanitize@url [0]{\catcode `\\12\catcode `\$12\catcode
  `\&12\catcode `\#12\catcode `\^12\catcode `\_12\catcode `\%12\relax}%
\providecommand \@@startlink[1]{}%
\providecommand \@@endlink[0]{}%
\providecommand \url  [0]{\begingroup\@sanitize@url \@url }%
\providecommand \@url [1]{\endgroup\@href {#1}{\urlprefix }}%
\providecommand \urlprefix  [0]{URL }%
\providecommand \Eprint [0]{\href }%
\providecommand \doibase [0]{http://dx.doi.org/}%
\providecommand \selectlanguage [0]{\@gobble}%
\providecommand \bibinfo  [0]{\@secondoftwo}%
\providecommand \bibfield  [0]{\@secondoftwo}%
\providecommand \translation [1]{[#1]}%
\providecommand \BibitemOpen [0]{}%
\providecommand \bibitemStop [0]{}%
\providecommand \bibitemNoStop [0]{.\EOS\space}%
\providecommand \EOS [0]{\spacefactor3000\relax}%
\providecommand \BibitemShut  [1]{\csname bibitem#1\endcsname}%
\let\auto@bib@innerbib\@empty
\bibitem [{\citenamefont {Li}\ \emph {et~al.}(2019)\citenamefont {Li},
  \citenamefont {Lee}, \citenamefont {Wang}, \citenamefont {Osada},
  \citenamefont {Crossley}, \citenamefont {Lee}, \citenamefont {Cui},
  \citenamefont {Hikita},\ and\ \citenamefont {Hwang}}]{li-nature19}%
  \BibitemOpen
  \bibfield  {author} {\bibinfo {author} {\bibfnamefont {D.}~\bibnamefont
  {Li}}, \bibinfo {author} {\bibfnamefont {K.}~\bibnamefont {Lee}}, \bibinfo
  {author} {\bibfnamefont {B.~Y.}\ \bibnamefont {Wang}}, \bibinfo {author}
  {\bibfnamefont {M.}~\bibnamefont {Osada}}, \bibinfo {author} {\bibfnamefont
  {S.}~\bibnamefont {Crossley}}, \bibinfo {author} {\bibfnamefont {H.~R.}\
  \bibnamefont {Lee}}, \bibinfo {author} {\bibfnamefont {Y.}~\bibnamefont
  {Cui}}, \bibinfo {author} {\bibfnamefont {Y.}~\bibnamefont {Hikita}}, \ and\
  \bibinfo {author} {\bibfnamefont {H.~Y.}\ \bibnamefont {Hwang}},\ }\href
  {\doibase 10.1038/s41586-019-1496-5} {\bibfield  {journal} {\bibinfo
  {journal} {Nature}\ }\textbf {\bibinfo {volume} {572}},\ \bibinfo {pages}
  {624} (\bibinfo {year} {2019})}\BibitemShut {NoStop}%
\bibitem [{\citenamefont {Botana}\ and\ \citenamefont {Norman}()}]{botana19}%
  \BibitemOpen
  \bibfield  {author} {\bibinfo {author} {\bibfnamefont {A.~S.}\ \bibnamefont
  {Botana}}\ and\ \bibinfo {author} {\bibfnamefont {M.~R.}\ \bibnamefont
  {Norman}},\ }\href@noop {} {\ }\Eprint {http://arxiv.org/abs/1908.10946}
  {arXiv:1908.10946} \BibitemShut {NoStop}%
\bibitem [{\citenamefont {Sakakibara}\ \emph {et~al.}()\citenamefont
  {Sakakibara}, \citenamefont {Usui}, \citenamefont {Suzuki}, \citenamefont
  {Kotani}, \citenamefont {Aoki},\ and\ \citenamefont {Kuroki}}]{sakakibara19}%
  \BibitemOpen
  \bibfield  {author} {\bibinfo {author} {\bibfnamefont {H.}~\bibnamefont
  {Sakakibara}}, \bibinfo {author} {\bibfnamefont {H.}~\bibnamefont {Usui}},
  \bibinfo {author} {\bibfnamefont {K.}~\bibnamefont {Suzuki}}, \bibinfo
  {author} {\bibfnamefont {T.}~\bibnamefont {Kotani}}, \bibinfo {author}
  {\bibfnamefont {H.}~\bibnamefont {Aoki}}, \ and\ \bibinfo {author}
  {\bibfnamefont {K.}~\bibnamefont {Kuroki}},\ }\href@noop {} {\ }\Eprint
  {http://arxiv.org/abs/1909.00060} {arXiv:1909.00060} \BibitemShut {NoStop}%
\bibitem [{\citenamefont {{Hirsch}}\ and\ \citenamefont
  {{Marsiglio}}(2019)}]{hirsch19}%
  \BibitemOpen
  \bibfield  {author} {\bibinfo {author} {\bibfnamefont {J.~E.}\ \bibnamefont
  {{Hirsch}}}\ and\ \bibinfo {author} {\bibfnamefont {F.}~\bibnamefont
  {{Marsiglio}}},\ }\href {\doibase 10.1016/j.physc.2019.1353534} {\bibfield
  {journal} {\bibinfo  {journal} {Physica C Superconductivity}\ }\textbf
  {\bibinfo {volume} {566}},\ \bibinfo {pages} {1353534} (\bibinfo {year}
  {2019})}\BibitemShut {NoStop}%
\bibitem [{\citenamefont {{Jiang}}\ \emph {et~al.}()\citenamefont {{Jiang}},
  \citenamefont {{Berciu}},\ and\ \citenamefont {{Sawatzky}}}]{jiang19}%
  \BibitemOpen
  \bibfield  {author} {\bibinfo {author} {\bibfnamefont {M.}~\bibnamefont
  {{Jiang}}}, \bibinfo {author} {\bibfnamefont {M.}~\bibnamefont {{Berciu}}}, \
  and\ \bibinfo {author} {\bibfnamefont {G.~A.}\ \bibnamefont {{Sawatzky}}},\
  }\href@noop {} {\ }\Eprint {http://arxiv.org/abs/1909.02557}
  {arXiv:1909.02557} \BibitemShut {NoStop}%
\bibitem [{\citenamefont {Wu}\ \emph {et~al.}()\citenamefont {Wu},
  \citenamefont {Sante}, \citenamefont {Schwemmer}, \citenamefont {Hanke},
  \citenamefont {Hwang}, \citenamefont {Raghu},\ and\ \citenamefont
  {Thomale}}]{wu19}%
  \BibitemOpen
  \bibfield  {author} {\bibinfo {author} {\bibfnamefont {X.}~\bibnamefont
  {Wu}}, \bibinfo {author} {\bibfnamefont {D.~D.}\ \bibnamefont {Sante}},
  \bibinfo {author} {\bibfnamefont {T.}~\bibnamefont {Schwemmer}}, \bibinfo
  {author} {\bibfnamefont {W.}~\bibnamefont {Hanke}}, \bibinfo {author}
  {\bibfnamefont {H.~Y.}\ \bibnamefont {Hwang}}, \bibinfo {author}
  {\bibfnamefont {S.}~\bibnamefont {Raghu}}, \ and\ \bibinfo {author}
  {\bibfnamefont {R.}~\bibnamefont {Thomale}},\ }\href@noop {} {\ }\Eprint
  {http://arxiv.org/abs/1909.03015} {arXiv:1909.03015} \BibitemShut {NoStop}%
\bibitem [{\citenamefont {{Nomura}}\ \emph {et~al.}()\citenamefont {{Nomura}},
  \citenamefont {{Hirayama}}, \citenamefont {{Tadano}}, \citenamefont
  {{Yoshimoto}}, \citenamefont {{Nakamura}},\ and\ \citenamefont
  {{Arita}}}]{nomura19}%
  \BibitemOpen
  \bibfield  {author} {\bibinfo {author} {\bibfnamefont {Y.}~\bibnamefont
  {{Nomura}}}, \bibinfo {author} {\bibfnamefont {M.}~\bibnamefont
  {{Hirayama}}}, \bibinfo {author} {\bibfnamefont {T.}~\bibnamefont
  {{Tadano}}}, \bibinfo {author} {\bibfnamefont {Y.}~\bibnamefont
  {{Yoshimoto}}}, \bibinfo {author} {\bibfnamefont {K.}~\bibnamefont
  {{Nakamura}}}, \ and\ \bibinfo {author} {\bibfnamefont {R.}~\bibnamefont
  {{Arita}}},\ }\href@noop {} {\ }\Eprint {http://arxiv.org/abs/1909.03942}
  {arXiv:1909.03942} \BibitemShut {NoStop}%
\bibitem [{\citenamefont {Bernardini}\ \emph {et~al.}()\citenamefont
  {Bernardini}, \citenamefont {Olevano},\ and\ \citenamefont
  {Cano}}]{bernardini19}%
  \BibitemOpen
  \bibfield  {author} {\bibinfo {author} {\bibfnamefont {F.}~\bibnamefont
  {Bernardini}}, \bibinfo {author} {\bibfnamefont {V.}~\bibnamefont {Olevano}},
  \ and\ \bibinfo {author} {\bibfnamefont {A.}~\bibnamefont {Cano}},\
  }\href@noop {} {\ }\Eprint {http://arxiv.org/abs/1910.13269}
  {arXiv:1910.13269} \BibitemShut {NoStop}%
\bibitem [{\citenamefont {{Hepting {\it et al.}}}()}]{hepting19}%
  \BibitemOpen
  \bibfield  {author} {\bibinfo {author} {\bibfnamefont {M.}~\bibnamefont
  {{Hepting {\it et al.}}}},\ }\href@noop {} {\ }\Eprint
  {http://arxiv.org/abs/1909.02678} {arXiv:1909.02678} \BibitemShut {NoStop}%
\bibitem [{\citenamefont {Zhang}\ \emph {et~al.}()\citenamefont {Zhang},
  \citenamefont {Jin}, \citenamefont {Wang}, \citenamefont {Xi}, \citenamefont
  {Shi}, \citenamefont {Ye},\ and\ \citenamefont {Mei}}]{zhang19effective}%
  \BibitemOpen
  \bibfield  {author} {\bibinfo {author} {\bibfnamefont {H.}~\bibnamefont
  {Zhang}}, \bibinfo {author} {\bibfnamefont {L.}~\bibnamefont {Jin}}, \bibinfo
  {author} {\bibfnamefont {S.}~\bibnamefont {Wang}}, \bibinfo {author}
  {\bibfnamefont {B.}~\bibnamefont {Xi}}, \bibinfo {author} {\bibfnamefont
  {X.}~\bibnamefont {Shi}}, \bibinfo {author} {\bibfnamefont {F.}~\bibnamefont
  {Ye}}, \ and\ \bibinfo {author} {\bibfnamefont {J.-W.}\ \bibnamefont {Mei}},\
  }\href@noop {} {\ }\Eprint {http://arxiv.org/abs/1909.07427}
  {arXiv:1909.07427} \BibitemShut {NoStop}%
\bibitem [{\citenamefont {{Zhang}}\ and\ \citenamefont
  {{Vishwanath}}()}]{zhang19}%
  \BibitemOpen
  \bibfield  {author} {\bibinfo {author} {\bibfnamefont {Y.-H.}\ \bibnamefont
  {{Zhang}}}\ and\ \bibinfo {author} {\bibfnamefont {A.}~\bibnamefont
  {{Vishwanath}}},\ }\href@noop {} {\ }\Eprint
  {http://arxiv.org/abs/1909.12865} {arXiv:1909.12865} \BibitemShut {NoStop}%
\bibitem [{\citenamefont {{Zhang}}\ \emph {et~al.}()\citenamefont {{Zhang}},
  \citenamefont {{Yang}},\ and\ \citenamefont {{Zhang}}}]{zhang19a}%
  \BibitemOpen
  \bibfield  {author} {\bibinfo {author} {\bibfnamefont {G.-M.}\ \bibnamefont
  {{Zhang}}}, \bibinfo {author} {\bibfnamefont {Y.-F.}\ \bibnamefont {{Yang}}},
  \ and\ \bibinfo {author} {\bibfnamefont {F.-C.}\ \bibnamefont {{Zhang}}},\
  }\href@noop {} {\ }\Eprint {http://arxiv.org/abs/1909.11845}
  {arXiv:1909.11845} \BibitemShut {NoStop}%
\bibitem [{\citenamefont {Choi}\ \emph {et~al.}()\citenamefont {Choi},
  \citenamefont {Lee},\ and\ \citenamefont {Pickett}}]{pickett19}%
  \BibitemOpen
  \bibfield  {author} {\bibinfo {author} {\bibfnamefont {M.-Y.}\ \bibnamefont
  {Choi}}, \bibinfo {author} {\bibfnamefont {K.~W.}\ \bibnamefont {Lee}}, \
  and\ \bibinfo {author} {\bibfnamefont {W.~E.}\ \bibnamefont {Pickett}},\
  }\href@noop {} {\ }\Eprint {http://arxiv.org/abs/1911.02999}
  {arXiv:1911.02999} \BibitemShut {NoStop}%
\bibitem [{\citenamefont {{Hu}}\ and\ \citenamefont {{Wu}}()}]{hu19}%
  \BibitemOpen
  \bibfield  {author} {\bibinfo {author} {\bibfnamefont {L.-H.}\ \bibnamefont
  {{Hu}}}\ and\ \bibinfo {author} {\bibfnamefont {C.}~\bibnamefont {{Wu}}},\
  }\href@noop {} {\ }\Eprint {http://arxiv.org/abs/1910.02482}
  {arXiv:1910.02482} \BibitemShut {NoStop}%
\bibitem [{\citenamefont {{Ryee}}\ \emph {et~al.}(2019)\citenamefont {{Ryee}},
  \citenamefont {{Yoon}}, \citenamefont {{Kim}}, \citenamefont {{Jeong}},\ and\
  \citenamefont {{Han}}}]{ryee19}%
  \BibitemOpen
  \bibfield  {author} {\bibinfo {author} {\bibfnamefont {S.}~\bibnamefont
  {{Ryee}}}, \bibinfo {author} {\bibfnamefont {H.}~\bibnamefont {{Yoon}}},
  \bibinfo {author} {\bibfnamefont {T.~J.}\ \bibnamefont {{Kim}}}, \bibinfo
  {author} {\bibfnamefont {M.~Y.}\ \bibnamefont {{Jeong}}}, \ and\ \bibinfo
  {author} {\bibfnamefont {M.~J.}\ \bibnamefont {{Han}}},\ }\href@noop {}
  {\bibfield  {journal} {\bibinfo  {journal} {arXiv e-prints}\ ,\ \bibinfo
  {eid} {arXiv:1909.05824}} (\bibinfo {year} {2019})},\ \Eprint
  {http://arxiv.org/abs/1909.05824} {arXiv:1909.05824 [cond-mat.supr-con]}
  \BibitemShut {NoStop}%
\bibitem [{\citenamefont {{Hirayama}}\ \emph {et~al.}()\citenamefont
  {{Hirayama}}, \citenamefont {{Tadano}}, \citenamefont {{Nomura}},\ and\
  \citenamefont {{Arita}}}]{hirayama19}%
  \BibitemOpen
  \bibfield  {author} {\bibinfo {author} {\bibfnamefont {M.}~\bibnamefont
  {{Hirayama}}}, \bibinfo {author} {\bibfnamefont {T.}~\bibnamefont
  {{Tadano}}}, \bibinfo {author} {\bibfnamefont {Y.}~\bibnamefont {{Nomura}}},
  \ and\ \bibinfo {author} {\bibfnamefont {R.}~\bibnamefont {{Arita}}},\
  }\href@noop {} {\ }\Eprint {http://arxiv.org/abs/1910.03974}
  {arXiv:1910.03974} \BibitemShut {NoStop}%
\bibitem [{\citenamefont {Anisimov}\ \emph {et~al.}(1999)\citenamefont
  {Anisimov}, \citenamefont {Bukhvalov},\ and\ \citenamefont
  {Rice}}]{anisimov99}%
  \BibitemOpen
  \bibfield  {author} {\bibinfo {author} {\bibfnamefont {V.~I.}\ \bibnamefont
  {Anisimov}}, \bibinfo {author} {\bibfnamefont {D.}~\bibnamefont {Bukhvalov}},
  \ and\ \bibinfo {author} {\bibfnamefont {T.~M.}\ \bibnamefont {Rice}},\
  }\href {\doibase 10.1103/PhysRevB.59.7901} {\bibfield  {journal} {\bibinfo
  {journal} {Phys. Rev. B}\ }\textbf {\bibinfo {volume} {59}},\ \bibinfo
  {pages} {7901} (\bibinfo {year} {1999})}\BibitemShut {NoStop}%
\bibitem [{\citenamefont {Lee}\ and\ \citenamefont
  {Pickett}(2004)}]{pickett04}%
  \BibitemOpen
  \bibfield  {author} {\bibinfo {author} {\bibfnamefont {K.-W.}\ \bibnamefont
  {Lee}}\ and\ \bibinfo {author} {\bibfnamefont {W.~E.}\ \bibnamefont
  {Pickett}},\ }\href {\doibase 10.1103/PhysRevB.70.165109} {\bibfield
  {journal} {\bibinfo  {journal} {Phys. Rev. B}\ }\textbf {\bibinfo {volume}
  {70}},\ \bibinfo {pages} {165109} (\bibinfo {year} {2004})}\BibitemShut
  {NoStop}%
\bibitem [{\citenamefont {Crespin}\ \emph {et~al.}(1983)\citenamefont
  {Crespin}, \citenamefont {Levitz},\ and\ \citenamefont
  {Gatineau}}]{crespin83}%
  \BibitemOpen
  \bibfield  {author} {\bibinfo {author} {\bibfnamefont {M.}~\bibnamefont
  {Crespin}}, \bibinfo {author} {\bibfnamefont {P.}~\bibnamefont {Levitz}}, \
  and\ \bibinfo {author} {\bibfnamefont {L.}~\bibnamefont {Gatineau}},\ }\href
  {\doibase 10.1039/F29837901181} {\bibfield  {journal} {\bibinfo  {journal}
  {J. Chem. Soc.{,} Faraday Trans. 2}\ }\textbf {\bibinfo {volume} {79}},\
  \bibinfo {pages} {1181} (\bibinfo {year} {1983})}\BibitemShut {NoStop}%
\bibitem [{\citenamefont {Hayward}\ \emph {et~al.}(1999)\citenamefont
  {Hayward}, \citenamefont {Green}, \citenamefont {Rosseinsky},\ and\
  \citenamefont {Sloan}}]{hayward99}%
  \BibitemOpen
  \bibfield  {author} {\bibinfo {author} {\bibfnamefont {M.~A.}\ \bibnamefont
  {Hayward}}, \bibinfo {author} {\bibfnamefont {M.~A.}\ \bibnamefont {Green}},
  \bibinfo {author} {\bibfnamefont {M.~J.}\ \bibnamefont {Rosseinsky}}, \ and\
  \bibinfo {author} {\bibfnamefont {J.}~\bibnamefont {Sloan}},\ }\href
  {\doibase 10.1021/ja991573i} {\bibfield  {journal} {\bibinfo  {journal}
  {Journal of the American Chemical Society}\ }\textbf {\bibinfo {volume}
  {121}},\ \bibinfo {pages} {8843} (\bibinfo {year} {1999})},\ \Eprint
  {http://arxiv.org/abs/https://doi.org/10.1021/ja991573i}
  {https://doi.org/10.1021/ja991573i} \BibitemShut {NoStop}%
\bibitem [{\citenamefont {Crespin}\ \emph {et~al.}(2005)\citenamefont
  {Crespin}, \citenamefont {Isnard}, \citenamefont {Dubois}, \citenamefont
  {Choisnet},\ and\ \citenamefont {Odier}}]{crespin05}%
  \BibitemOpen
  \bibfield  {author} {\bibinfo {author} {\bibfnamefont {M.}~\bibnamefont
  {Crespin}}, \bibinfo {author} {\bibfnamefont {O.}~\bibnamefont {Isnard}},
  \bibinfo {author} {\bibfnamefont {F.}~\bibnamefont {Dubois}}, \bibinfo
  {author} {\bibfnamefont {J.}~\bibnamefont {Choisnet}}, \ and\ \bibinfo
  {author} {\bibfnamefont {P.}~\bibnamefont {Odier}},\ }\href {\doibase
  https://doi.org/10.1016/j.jssc.2005.01.023} {\bibfield  {journal} {\bibinfo
  {journal} {Journal of Solid State Chemistry}\ }\textbf {\bibinfo {volume}
  {178}},\ \bibinfo {pages} {1326 } (\bibinfo {year} {2005})}\BibitemShut
  {NoStop}%
\bibitem [{\citenamefont {Kresse}\ and\ \citenamefont
  {Furthmuller}(1996)}]{VASP}%
  \BibitemOpen
  \bibfield  {author} {\bibinfo {author} {\bibfnamefont {G.}~\bibnamefont
  {Kresse}}\ and\ \bibinfo {author} {\bibfnamefont {J.}~\bibnamefont
  {Furthmuller}},\ }\href {\doibase
  https://doi.org/10.1016/0927-0256(96)00008-0} {\bibfield  {journal} {\bibinfo
   {journal} {Computational Materials Science}\ }\textbf {\bibinfo {volume}
  {6}},\ \bibinfo {pages} {15 } (\bibinfo {year} {1996})}\BibitemShut {NoStop}%
\bibitem [{\citenamefont {Kresse}\ and\ \citenamefont {Joubert}(1999)}]{PAW}%
  \BibitemOpen
  \bibfield  {author} {\bibinfo {author} {\bibfnamefont {G.}~\bibnamefont
  {Kresse}}\ and\ \bibinfo {author} {\bibfnamefont {D.}~\bibnamefont
  {Joubert}},\ }\href {\doibase 10.1103/PhysRevB.59.1758} {\bibfield  {journal}
  {\bibinfo  {journal} {Phys. Rev. B}\ }\textbf {\bibinfo {volume} {59}},\
  \bibinfo {pages} {1758} (\bibinfo {year} {1999})}\BibitemShut {NoStop}%
\bibitem [{\citenamefont {Perdew}\ \emph {et~al.}(1996)\citenamefont {Perdew},
  \citenamefont {Burke},\ and\ \citenamefont {Ernzerhof}}]{PBE}%
  \BibitemOpen
  \bibfield  {author} {\bibinfo {author} {\bibfnamefont {J.~P.}\ \bibnamefont
  {Perdew}}, \bibinfo {author} {\bibfnamefont {K.}~\bibnamefont {Burke}}, \
  and\ \bibinfo {author} {\bibfnamefont {M.}~\bibnamefont {Ernzerhof}},\ }\href
  {\doibase 10.1103/PhysRevLett.77.3865} {\bibfield  {journal} {\bibinfo
  {journal} {Phys. Rev. Lett.}\ }\textbf {\bibinfo {volume} {77}},\ \bibinfo
  {pages} {3865} (\bibinfo {year} {1996})}\BibitemShut {NoStop}%
\bibitem [{\citenamefont {Blaha}\ \emph {et~al.}()\citenamefont {Blaha},
  \citenamefont {Schwarz}, \citenamefont {Madsen}, \citenamefont {Kvasnicka},
  \citenamefont {Luitz}, \citenamefont {Laskowski}, \citenamefont {Tran},\ and\
  \citenamefont {Marks}}]{Wien2k}%
  \BibitemOpen
  \bibfield  {author} {\bibinfo {author} {\bibfnamefont {P.}~\bibnamefont
  {Blaha}}, \bibinfo {author} {\bibfnamefont {K.}~\bibnamefont {Schwarz}},
  \bibinfo {author} {\bibfnamefont {G.}~\bibnamefont {Madsen}}, \bibinfo
  {author} {\bibfnamefont {D.}~\bibnamefont {Kvasnicka}}, \bibinfo {author}
  {\bibfnamefont {J.}~\bibnamefont {Luitz}}, \bibinfo {author} {\bibfnamefont
  {R.}~\bibnamefont {Laskowski}}, \bibinfo {author} {\bibfnamefont
  {F.}~\bibnamefont {Tran}}, \ and\ \bibinfo {author} {\bibfnamefont {L.~D.}\
  \bibnamefont {Marks}},\ }\href@noop {} {\bibinfo  {journal} {{W}{I}{E}{N}2k,
  An Augmented Plane Wave + Local Orbitals Program for Calculating Crystal
  Properties (Karlheinz Schwarz, Techn. Universität Wien, Austria), 2018. ISBN
  3-9501031-1-2}\ }\BibitemShut {NoStop}%
\bibitem [{\citenamefont {Perdew}\ and\ \citenamefont {Zunger}(1981)}]{LDA}%
  \BibitemOpen
\bibfield  {journal} {  }\bibfield  {author} {\bibinfo {author} {\bibfnamefont
  {J.~P.}\ \bibnamefont {Perdew}}\ and\ \bibinfo {author} {\bibfnamefont
  {A.}~\bibnamefont {Zunger}},\ }\href {\doibase 10.1103/PhysRevB.23.5048}
  {\bibfield  {journal} {\bibinfo  {journal} {Phys. Rev. B}\ }\textbf {\bibinfo
  {volume} {23}},\ \bibinfo {pages} {5048} (\bibinfo {year}
  {1981})}\BibitemShut {NoStop}%
\bibitem [{\citenamefont {Marzari}\ \emph {et~al.}(2012)\citenamefont
  {Marzari}, \citenamefont {Mostofi}, \citenamefont {Yates}, \citenamefont
  {Souza},\ and\ \citenamefont {Vanderbilt}}]{MLWF}%
  \BibitemOpen
  \bibfield  {author} {\bibinfo {author} {\bibfnamefont {N.}~\bibnamefont
  {Marzari}}, \bibinfo {author} {\bibfnamefont {A.~A.}\ \bibnamefont
  {Mostofi}}, \bibinfo {author} {\bibfnamefont {J.~R.}\ \bibnamefont {Yates}},
  \bibinfo {author} {\bibfnamefont {I.}~\bibnamefont {Souza}}, \ and\ \bibinfo
  {author} {\bibfnamefont {D.}~\bibnamefont {Vanderbilt}},\ }\href {\doibase
  10.1103/RevModPhys.84.1419} {\bibfield  {journal} {\bibinfo  {journal} {Rev.
  Mod. Phys.}\ }\textbf {\bibinfo {volume} {84}},\ \bibinfo {pages} {1419}
  (\bibinfo {year} {2012})}\BibitemShut {NoStop}%
\bibitem [{\citenamefont {Mostofi}\ \emph {et~al.}(2008)\citenamefont
  {Mostofi}, \citenamefont {Yates}, \citenamefont {Lee}, \citenamefont {Souza},
  \citenamefont {Vanderbilt},\ and\ \citenamefont {Marzari}}]{Wannier90}%
  \BibitemOpen
  \bibfield  {author} {\bibinfo {author} {\bibfnamefont {A.~A.}\ \bibnamefont
  {Mostofi}}, \bibinfo {author} {\bibfnamefont {J.~R.}\ \bibnamefont {Yates}},
  \bibinfo {author} {\bibfnamefont {Y.-S.}\ \bibnamefont {Lee}}, \bibinfo
  {author} {\bibfnamefont {I.}~\bibnamefont {Souza}}, \bibinfo {author}
  {\bibfnamefont {D.}~\bibnamefont {Vanderbilt}}, \ and\ \bibinfo {author}
  {\bibfnamefont {N.}~\bibnamefont {Marzari}},\ }\href {\doibase
  https://doi.org/10.1016/j.cpc.2007.11.016} {\bibfield  {journal} {\bibinfo
  {journal} {Computer Physics Communications}\ }\textbf {\bibinfo {volume}
  {178}},\ \bibinfo {pages} {685 } (\bibinfo {year} {2008})}\BibitemShut
  {NoStop}%
\bibitem [{\citenamefont {Kunes}\ \emph {et~al.}(2010)\citenamefont {Kunes},
  \citenamefont {Arita}, \citenamefont {Wissgott}, \citenamefont {Toschi},
  \citenamefont {Ikeda},\ and\ \citenamefont {Held}}]{Wien2wannier}%
  \BibitemOpen
  \bibfield  {author} {\bibinfo {author} {\bibfnamefont {J.}~\bibnamefont
  {Kunes}}, \bibinfo {author} {\bibfnamefont {R.}~\bibnamefont {Arita}},
  \bibinfo {author} {\bibfnamefont {P.}~\bibnamefont {Wissgott}}, \bibinfo
  {author} {\bibfnamefont {A.}~\bibnamefont {Toschi}}, \bibinfo {author}
  {\bibfnamefont {H.}~\bibnamefont {Ikeda}}, \ and\ \bibinfo {author}
  {\bibfnamefont {K.}~\bibnamefont {Held}},\ }\href {\doibase
  https://doi.org/10.1016/j.cpc.2010.08.005} {\bibfield  {journal} {\bibinfo
  {journal} {Computer Physics Communications}\ }\textbf {\bibinfo {volume}
  {181}},\ \bibinfo {pages} {1888 } (\bibinfo {year} {2010})}\BibitemShut
  {NoStop}%
\bibitem [{\citenamefont {Bousquet}\ and\ \citenamefont
  {Cano}(2016)}]{bousquet16}%
  \BibitemOpen
  \bibfield  {author} {\bibinfo {author} {\bibfnamefont {E.}~\bibnamefont
  {Bousquet}}\ and\ \bibinfo {author} {\bibfnamefont {A.}~\bibnamefont
  {Cano}},\ }\href {\doibase 10.1088/0953-8984/28/12/123001} {\bibfield
  {journal} {\bibinfo  {journal} {J. Phys. Condens. Matter}\ }\textbf {\bibinfo
  {volume} {28}},\ \bibinfo {pages} {123001} (\bibinfo {year}
  {2016})}\BibitemShut {NoStop}%
\bibitem [{\citenamefont {Garcia-Castro}\ \emph {et~al.}(2014)\citenamefont
  {Garcia-Castro}, \citenamefont {Spaldin}, \citenamefont {Romero},\ and\
  \citenamefont {Bousquet}}]{garcia-castro-prb13}%
  \BibitemOpen
  \bibfield  {author} {\bibinfo {author} {\bibfnamefont {A.~C.}\ \bibnamefont
  {Garcia-Castro}}, \bibinfo {author} {\bibfnamefont {N.~A.}\ \bibnamefont
  {Spaldin}}, \bibinfo {author} {\bibfnamefont {A.~H.}\ \bibnamefont {Romero}},
  \ and\ \bibinfo {author} {\bibfnamefont {E.}~\bibnamefont {Bousquet}},\
  }\href {\doibase 10.1103/PhysRevB.89.104107} {\bibfield  {journal} {\bibinfo
  {journal} {Phys. Rev. B}\ }\textbf {\bibinfo {volume} {89}},\ \bibinfo
  {pages} {104107} (\bibinfo {year} {2014})}\BibitemShut {NoStop}%
\bibitem [{\citenamefont {Sakakibara}\ \emph {et~al.}(2012)\citenamefont
  {Sakakibara}, \citenamefont {Usui}, \citenamefont {Kuroki}, \citenamefont
  {Arita},\ and\ \citenamefont {Aoki}}]{sakakibara12}%
  \BibitemOpen
  \bibfield  {author} {\bibinfo {author} {\bibfnamefont {H.}~\bibnamefont
  {Sakakibara}}, \bibinfo {author} {\bibfnamefont {H.}~\bibnamefont {Usui}},
  \bibinfo {author} {\bibfnamefont {K.}~\bibnamefont {Kuroki}}, \bibinfo
  {author} {\bibfnamefont {R.}~\bibnamefont {Arita}}, \ and\ \bibinfo {author}
  {\bibfnamefont {H.}~\bibnamefont {Aoki}},\ }\href {\doibase
  10.1103/PhysRevB.85.064501} {\bibfield  {journal} {\bibinfo  {journal} {Phys.
  Rev. B}\ }\textbf {\bibinfo {volume} {85}},\ \bibinfo {pages} {064501}
  (\bibinfo {year} {2012})}\BibitemShut {NoStop}%
\bibitem [{\citenamefont {Jiang}\ \emph {et~al.}()\citenamefont {Jiang},
  \citenamefont {Si}, \citenamefont {Liao},\ and\ \citenamefont
  {Zhong}}]{jiang2019electronic}%
  \BibitemOpen
  \bibfield  {author} {\bibinfo {author} {\bibfnamefont {P.}~\bibnamefont
  {Jiang}}, \bibinfo {author} {\bibfnamefont {L.}~\bibnamefont {Si}}, \bibinfo
  {author} {\bibfnamefont {Z.}~\bibnamefont {Liao}}, \ and\ \bibinfo {author}
  {\bibfnamefont {Z.}~\bibnamefont {Zhong}},\ }\href@noop {} {\ }\Eprint
  {http://arxiv.org/abs/1909.13634} {arXiv:1909.13634} \BibitemShut {NoStop}%
\bibitem [{\citenamefont {{P. Villar Arribi}}\ \emph {et~al.}()\citenamefont
  {{P. Villar Arribi}}, \citenamefont {Bernardini}, \citenamefont {de'Medici},
  \citenamefont {Toulemonde}, \citenamefont {Tencé},\ and\ \citenamefont
  {Cano}}]{arribi2018}%
  \BibitemOpen
  \bibfield  {author} {\bibinfo {author} {\bibnamefont {{P. Villar Arribi}}},
  \bibinfo {author} {\bibfnamefont {F.}~\bibnamefont {Bernardini}}, \bibinfo
  {author} {\bibfnamefont {L.}~\bibnamefont {de'Medici}}, \bibinfo {author}
  {\bibfnamefont {P.}~\bibnamefont {Toulemonde}}, \bibinfo {author}
  {\bibfnamefont {S.}~\bibnamefont {Tencé}}, \ and\ \bibinfo {author}
  {\bibfnamefont {A.}~\bibnamefont {Cano}},\ }\href@noop {} {\ }\Eprint
  {http://arxiv.org/abs/1810.10306} {arXiv:1810.10306} \BibitemShut {NoStop}%
\end{thebibliography}%

\newpage 

\onecolumngrid

\section*{Appendix}

\end{document}